\documentclass[usegraphicx,useAMS,usenatbib]{mn2e}
\usepackage{times}
\usepackage{amssymb}
\usepackage{amsmath}
\usepackage{graphicx}
\usepackage{euscript}
\usepackage{rotating}
\usepackage{epsfig}
\usepackage{mathptmx}
\usepackage{amsbsy}

\def\arc{{\rm\thinspace arcsec}}
\def\cm{{\rm\thinspace cm}}

\def\erg{{\rm\thinspace erg}}

\def\ha{H$\alpha$}
\def\hb{H$\beta$}

\def\nii{[N{\sc ii}]$\lambda$6584}

\def\km{{\rm\thinspace km}}

\def\s{{\rm\thinspace s}}
\def\yr{{\rm\thinspace yr}}

\def\ergpcmsqpspsqarcsec{\hbox{$\erg\cm^{-2}\s^{-1}\arc^{-2}\,$}}

\def\kmps{\hbox{$\km\s^{-1}\,$}}

\def\ps{\hbox{$\s^{-1}\,$}}

\DeclareMathAlphabet{\vib}{OML}{cmm}{m}{it}
\begin{document}
\include{defn}
\title{On the origin and excitation of the extended nebula surrounding
  NGC\,1275} \author[N. A. Hatch, C. S. Crawford, A. C. Fabian \& R.
M. Johnstone]
{N. A. Hatch,\thanks{E-mail:nah@ast.cam.ac.uk} C. S. Crawford, R. M. Johnstone and A. C. Fabian\\
  Institute of Astronomy, Madingley Road, Cambridge, CB3 0HA}
\maketitle
\begin{abstract}
  We use line-of-sight velocity information on the filamentary
  emission-line nebula of NGC\,1275 to infer a dynamical model of the
  nebula's flow through the surrounding intracluster gas.  We detect
  outflowing gas and flow patterns that match simulations of buoyantly
  rising bubbles from which we deduce that some of the nebula
  filaments have been drawn out of NGC\,1275. We find a radial
  gradient of the ratio [N{\sc ii}]$\lambda$6584/\ha\ which may be due
  to a variation in metallicity, interactions with the surrounding
  intracluster medium or a hardening of the excitation mechanism. We
  find no preferred spatial correlation of stellar clusters within the
  filaments and there is a notable lack of [O{\sc iii}]$\lambda$5007
  emission, therefore it is unlikely that the filaments are ionized by
  stellar UV. 
\end{abstract}
\section{Introduction}
NGC\,1275 is the central galaxy of the X-ray luminous Perseus cluster
(A426), which has bright centrally peaked X-ray emission, and a cool
core with a central temperature of a third the virial temperature
\citep{Schmidt, Sanders}. Cavities in the X-ray emission are observed
in a number of locations surrounding the central galaxy
\citep{Fabian2003b}. Where these coincide with GHz radio emission they
are interpreted as bubbles filled with relativistic plasma, which has
been injected into the intracluster medium (ICM) by the central
engine.  Cavities with no observed radio emission have been described
as `ghost bubbles', and are thought to have originated from an earlier
epoch of activity in the central engine.

\citet{Minkowski} discovered that the nebula of NGC\,1275 comprises of
two distinct emission-line systems: a high-velocity system
(8200\,km\ps), identified as a disrupted foreground galaxy
\citep{Boroson} at least 60\,kpc in front of NGC\,1275
\citep{Gillmon}, and a low-velocity system (5265\,km\ps). Although the
high-velocity system lies directly in front of NGC\,1275, the emission
lines are easily distinguished in wavelength from those of the
low-velocity system, and clearly indicate photoionization by hot,
young stars \citep{Kent}.

The low-velocity system associated with the central galaxy NGC\,1275
is the focus of this work. It is known to extend over 100\,kpc in a
large array of thin filaments \citep{Lynds,conselice}.  Whilst the
nebula is extremely luminous--4.1$\times10^{42}$\erg\ps\ in \ha\ and
[N{\sc ii}] \citep{Heckman}, with a total line luminosity probably 20
times that in \ha-- the power source remains unknown.  Ionization by
the central active galactic nucleus (AGN) residing in NGC\,1275 can be
ruled out as the dominant source of power for the extended nebula on
the grounds that the \ha\ luminosity does not decrease with distance
from the nucleus \citep{RodAndy}, although it may be important in the
luminous inner regions.  Ionization by hot young stars is an
attractive option as it is a local mechanism, but the line ratios are
drastically different to those seen in H{\sc ii} regions \citep{Kent}.
Models of heating by X-rays from the ICM have been put forward
\citep{Donahue91}, as well as conduction from the ICM \citep{Donahue},
shocks \citep{SabraShieldsFlip} and turbulent mixing layers
\citep{Crawfordmixinglayers}.

Soft X-ray emission is associated with some of the optical filaments
of NGC\,1275 \citep{Fabian2003}. The filaments are less luminous in
the X-ray than the optical/UV by up to two orders of magnitude
implying they are not excited by X-radiation.  The soft X-ray emission
indicates an interaction between the warm filaments and the hot ICM,
possibly via heat conduction.

Large deposits of molecular hydrogen have been discovered in the
central regions of NGC\,1275 \citep{Krabbe,Donahue} similar to other
central cluster galaxies with emission-line nebulae \citep{Edge}.
Recently, molecular hydrogen was observed in the outer filaments of
NGC\,1275 \citep{Hatch}, indicating gas at 2000\,K exists within the
hot ICM at radii of over 25\,kpc.

The origin of the filaments remain a mystery. Current theories include
condensing gas from the ICM in the form of a cooling flow \citep{FNC,
  Heckman, Donahue91}, gas accreting from previous mergers
\citep{Braine}, the explosive expulsion of gas from NGC\,1275
\citep{Burbidge} or gas drawn out \citep{Fabian2003}. The filaments
are very thin, long and the majority are radial. Submerged within the
ICM, they enable us to constrain the level of turbulence in the ICM
and argue for a laminar flow. As the ICM moves it may drag the warm
optically emitting gas, thus the filaments can act as streamlines
tracing the flow direction\citep{Fabian2003}.

None of these problems are exclusive to NGC\,1275 as extended
emission-line nebulae are commonly found surrounding other massive
galaxies in the centre of X-ray bright `cool cores', where the X-ray
emission is centrally peaked \citep{Crawford1999}.

This work presents new spectroscopic data that explore the kinematic
and line-emission properties of the nebula that surrounds NGC\,1275.
After analysing and interpreting the kinematics we put forward a
dynamical model of the nebula and discuss the origin of the filaments.
The redshift of NGC 1275 is 0.0176, which using
H$_{0}$=70\,kms$^{-1}$Mpc$^{-1}$, gives 1\,kpc$\simeq$2.7\,arcsec.
\section{Observations and data reduction}

\begin{figure*}
  \centering
\includegraphics[width=2.0\columnwidth]{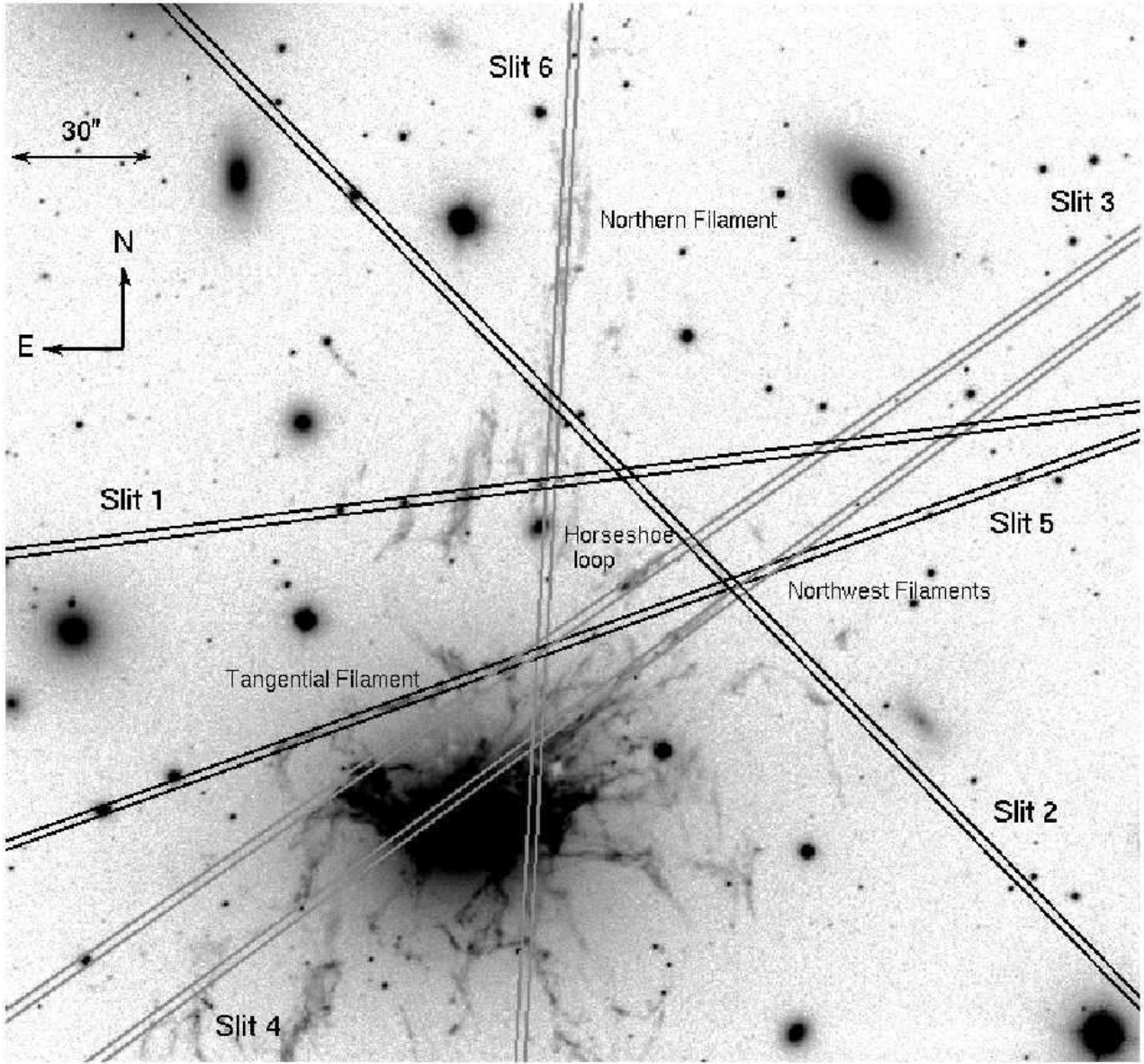}
\caption{Position of the six longslit observations on the \ha\ nebula
  surrounding NGC\,1275.  The background image is taken from the data
  of \citet{conselice}\label{slit_positions}}
\end{figure*}

The data were obtained on the nights of 2004 Sept 23 and 2004 Oct 06
using the GMOS North instrument on the Gemini North telescope on Mauna
Kea, Hawaii.  The sky on both nights was photometric and the seeing
was less than 0.8 arcsec. Six slit positions were chosen using the map
of Conselice et al. (2001; see Fig \ref{slit_positions}), selected to
feature particular structures of interest.  Two bright stars were
aligned on each slit so its exact position was known. The slit width
was 0.5 arcsec, filter R831+G590 was used and the exposure time for
each slit was 900 seconds (3$\times$300s); all exposures were binned
2$\times$2 before readout.  This setup allows us to explore the
observed wavelength range 4850--6945\AA.  The spectroscopic standards
BD+28d4211 and G1912B2 were observed in order to flux-calibrate the
data. A flat was taken with the GCAL instrument after every slit
position observed.  CuAr arcs and bias frames were taken with the
GCAL instrument on the nights of the 2004 Sept 23 and 2004 Oct 05. A
summary of all science exposures taken is given in Table
\ref{observations}.

The data were reduced using the IRAF Gemini package (version 1.7). The
bias frames were combined, flatfields were normalised and mosaiced
together. All science and standard star frames were reduced by bias
subtraction, mosaicing the individual chips together, flatfielding and
interpolating across the detector gaps. The arcs were calibrated and
checked manually against a line list before the wavelength solution
was transferred to the data. The science observations were flux
calibrated using the standard star G1912B2 (BD+28d4211 was not used as
it was observed 35$^{\circ}$ from the parallactic angle and showed
signs of strong atmospheric differential refraction), and finally
sky-subtracted. The data were de-reddened for Galactic extinction
using E(B--V)=0.315.  The spectra were extracted, converted to ASCII
format and analysis was performed using QDP \citep{qdp}. The red
section covering [O{\sc i}]$\lambda$6300 to [S{\sc ii}]$\lambda$6731
was fitted separately to a blue section covering the lines of \hb\ to
[N{\sc i}]$\lambda$5199. The He{\sc i}$\lambda$5876 was fitted
separately.  The lines were assumed to be Gaussian in profile and
share the same redshift and velocity width.  [N{\sc ii}]$\lambda$6548
was assumed to have a third of the intensity of the [N{\sc
  ii}]$\lambda$6584 line.

Acquisition images were taken at all slit positions through the
r\,G0303 filter. These images were reduced by bias subtraction,
flatfielding, then normalised and combined using the IRAF IMAGES package
to form an {\it R}-band image with a scale of 0.1454 arcsec/pixel. The
image was used to accurately align the slits with the \ha\ image from
\citet{conselice}.

\begin{table}
\centering
\begin{tabular}{||l|c|c||c|}\hline
Position & Exposure type& Filter& Exposure time (sec)\\\hline
Slit 1&Spectral&R831+G590&900\\ 
Slit 1&Image&rG0303&100\\ 
Slit 2&Spectral&R831+G590&900\\ 
Slit 3&Spectral&R831+G590&900\\ 
Slit 3&Image&rG0303&60\\ 
Slit 4&Spectral&R831+G590&900\\ 
Slit 4&Image&rG0303&300\\ 
Slit 5&Spectral&R831+G590&900\\ 
Slit 5&Image&rG0303&80\\ 
Slit 6&Spectral&R831+G590&900\\ 
Slit 6&Image&rG0303&300 \\ \hline
\end{tabular}
\caption{\label{observations} Summery of Gemini observations.}
\end{table}

Figure \ref{slit_positions} details all six longslit positions. Slit 1
cuts through the Northern filaments. Slit 2 covers the top of the
`horseshoe' feature to the Northwest and cuts through the main
Northern filament. Slit 3 covers the short straight radial western
part of the `horseshoe' feature and cuts across the North-East of the
nebula.  Slit 4 covers the long 30\,kpc radial filament that runs from
the galaxy into the `horseshoe' loop on the Eastern side.  Slit 5 runs
along the North of the nebula covering a long tangential filament.
Slit 6 covers the dominant Northern filament and passes through the
Western side of the nebula.

\section{Archival HST data}
\label{sec:HST}  

We extracted archival Hubble Space Telescope WFPC2 imaging data on
NGC\,1275 in three broad bands: F450W (with total exposure time of
5100\s), F702W (4700\s) and F814W (4800\s). At the redshift of
NGC\,1275, the strong emission lines of \ha+[N{\sc ii}] (from the
low-velocity system only) appear in the F702W bandpass.

\begin{figure*}
  \centering
  \includegraphics[width=1.8\columnwidth]{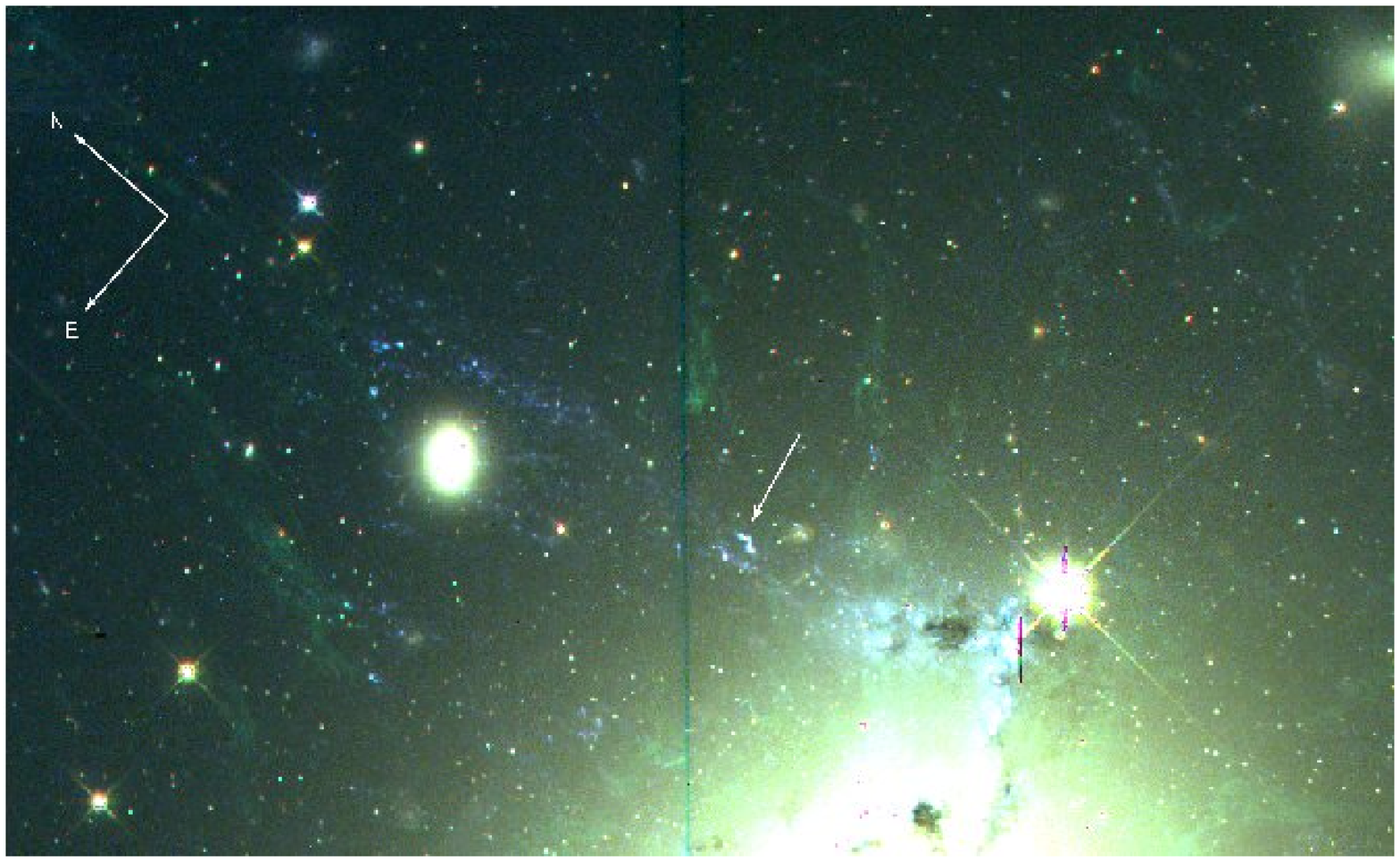} 
  \caption{Combined colour image of the North and West environment of
    NGC\,1275 as seen through the HST F450W (shown as blue), F702W
    (green) and F814W (red) broad-band filters.  The F702W filter
    encompasses \ha+[N{\sc ii}] line emission at the redshift of
    NGC\,1275. The image is 120 arcsec by 73 arcsec in size.  The
    Northwest and Northern filaments including the `horseshoe' loop
    are visible. A 1-arcminute-long chain of blue stellar clusters
    runs from the top left (above the cluster galaxy) toward the
    bottom right where it meets the Western edge of the infalling
    galaxy (high-velocity system) near the bright star. The white
    arrow points to a region where the detected \ha\ emission is
    redshifted by 5538\kmps\ placing it in the low-velocity system.  }
  \label{hstNW3col}
\end{figure*}

\begin{figure*}
  \centering
  \includegraphics[width=1.8\columnwidth]{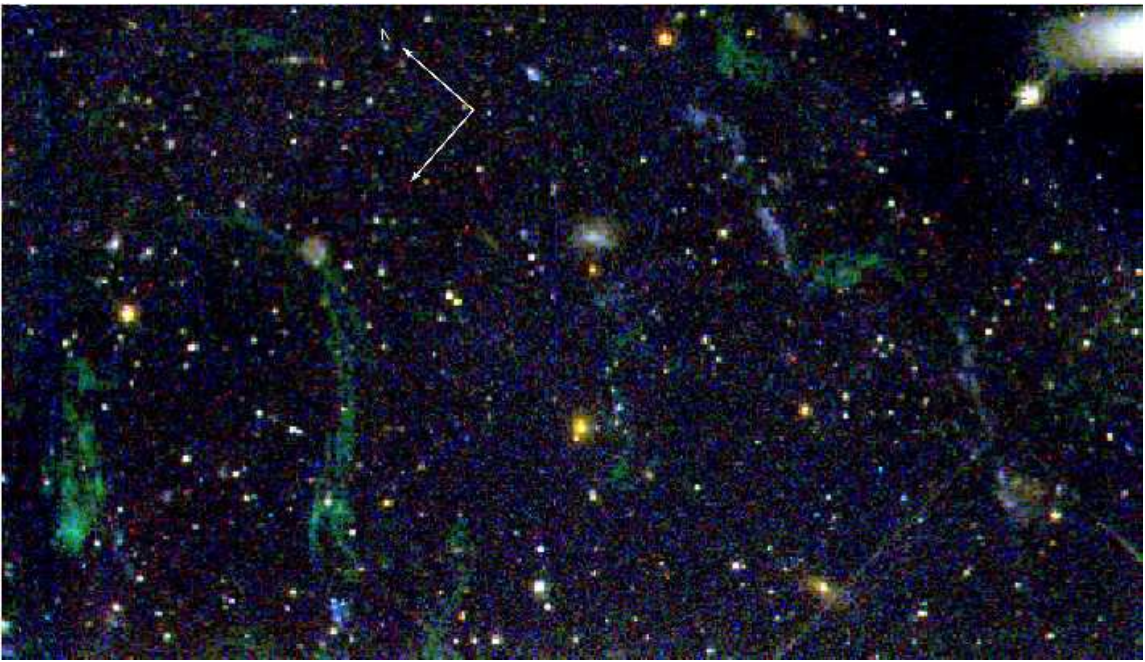} 
  \caption{A zoom to the Northwest of the previous figures, showing
detail of the loops in the region of the `horseshoe' filament. The three
colour images have each been unsharp-masked to remove the light from
the underlying central galaxy.   The image is 66 by 38 arcsec in size. 
    \label{hstloops}}
\end{figure*}

\begin{figure}
  \centering
    \includegraphics[width=1.0\columnwidth]{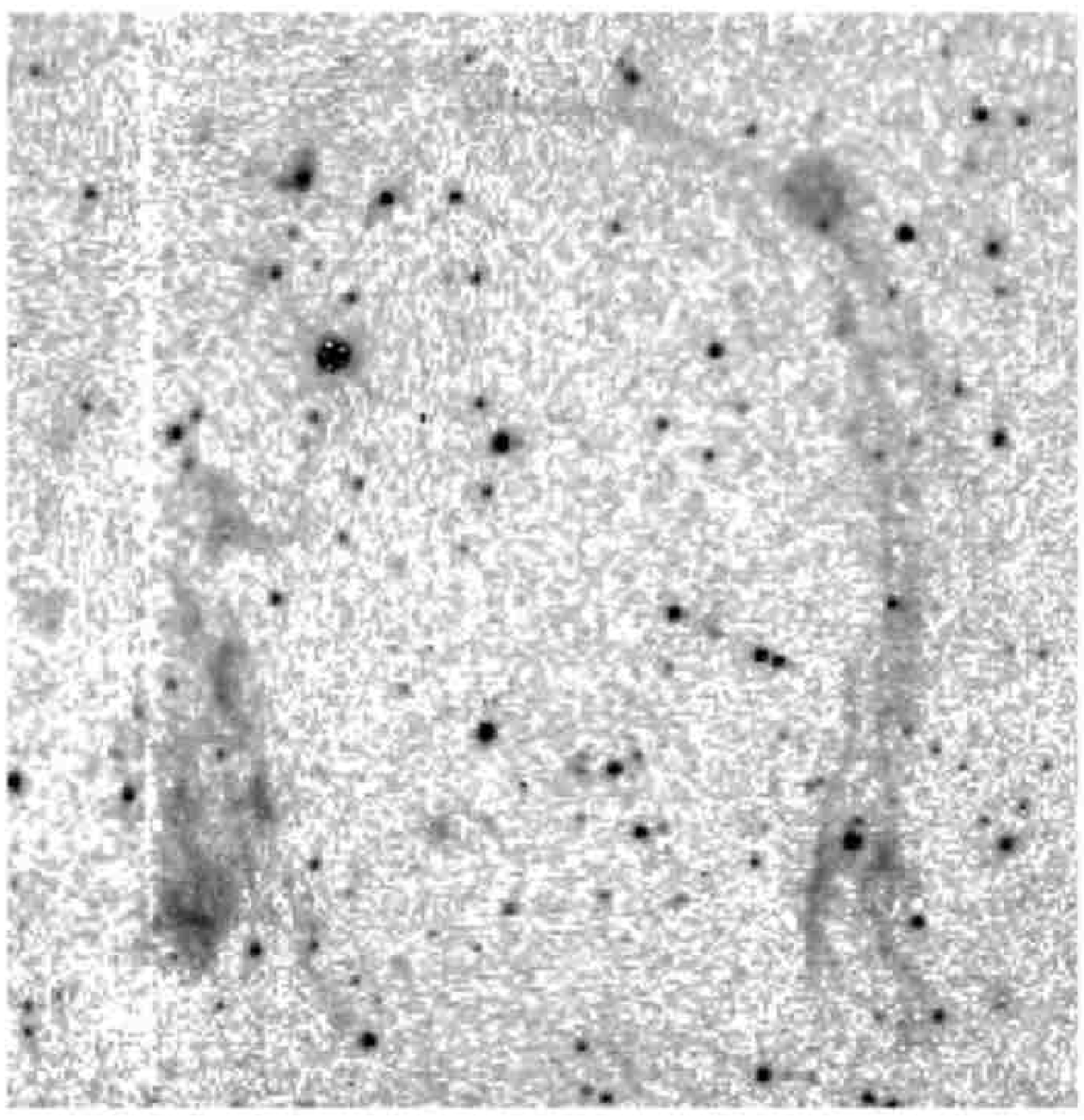} 
    \caption{A detailed image of the \ha\ \lq horseshoe' filament as
      seen through the HST F702W broad-band filter, showing the fine
      structure and smoothness of the individual strands.  The image
      is 24 arcsec on a side, and has been unsharp-masked (i.e. it has
      had a highly smoothed version of itself subtracted) to remove the
      underlying gradient due to the light from the central galaxy
      continuum.
    \label{hstfiladetail}}
\end{figure}

\section{Stellar clusters and continuum }
\label{sec:continuum} 
NGC\,1275 is surrounded by many star clusters
\citep{Holtzman,Carlson}, which are clearly visible strewn across the
HST image shown in Fig~\ref{hstNW3col}. There is a bimodal
distribution of colours, with the blue clusters more centrally located
\citep{Carlson} and some are possibly associated with the high-velocity
system. The majority of the stellar clusters appear red in colour, and
are scattered uniformly around the galaxy. Clusters are visible up to
the edge of the Gemini acquisition image, beyond 80\,kpc from
NGC\,1275. Although there are brighter, clumpy structures within the
line-emitting filaments, there does not appear to be any preferential
spatial association of the star clusters with the filaments in either
the Gemini or HST images (e.g. Figures~\ref{hstloops}, \ref{hstfiladetail}).

Figure \ref{hstNW3col} does, however, show a 1-arcminute-long chain of
exceptionally blue star clusters that stretch from just north of the
small cluster galaxy at RA 03:19:46.7, Dec 41:31:45.6 (J2000), down
into the western edge of the intervening high-velocity system.  This
chain is an extension of the blue clusters noted by \citet{conselice}
in their Figure 8. It is absent from the \ha\ image of the low
velocity system (Fig. \ref{slit_positions}) and from \ha\ images of
the intervening high velocity system \citep{Caulet}. The only blue
knot that coincides with \ha\ emission is marked by a white arrow in
Fig.~\ref{hstNW3col}, it has a line-of-sight velocity of 5538\kmps\
which firmly places it within the low velocity system. The spatial
connection and similarity in colour to the structures within the
high-velocity system suggest that many of these blue star clusters may
be associated with the disrupted foreground galaxy.  The
high-resolution HST image shows the \ha\ filaments themselves to be
highly smooth (Figure \ref{hstfiladetail}) and continuous, consisting
of several strands which are individually unresolved on scales of 0.25
arcsec (corresponding to 90 parsecs).

Excluding the central region where continuum from the galaxy is
prominent, our spectra show signs of continuum only at 9 regions. Two
of these regions exhibit no line emission so the continuum may have
originated from stellar clusters which typically do not have \ha\
emission \citep{Holtzman}. The other 7 regions exhibited line emission
with very similar line intensity ratios to regions without continuum
and only one exhibited [O{\sc iii}]$\lambda$5007 emission (see Figure
\ref{spectrumOIII} in section \ref{sec:OIII}).  The \ha\ luminosity
from these continuum regions is fairly typical. Although the continuum
regions are distinct in the R-band acquisition image as stellar
clusters, the HST images show that these bright knots are ubiquitous
throughout the region stretching beyond the emission-line nebula.
These are likely to be chance associations of alignment. It does not
appear that the stellar clusters are formed or located within the
filaments.

\section{Kinematics of the nebula}
\label{sec:kinematics}
The morphology of the filaments suggest they act as streamlines
tracing the gas flow in the ICM. Therefore the Doppler shifts of the
filaments may reveal the velocity field in the core of the cluster,
near NGC\,1275. The forces that could drive the flow of filaments are
gravity (from NGC\,1275), which would draw the filaments inwards, or
the outward pull following a buoyantly rising radio/ghost bubble as
proposed by \citet{BohringerM87,Churazov,Reynolds}. A three
dimensional flow pattern can differentiate between galactic outflow
and inflow models, and thus constrain the origin of the filaments. The
velocities are determined from binning the spatial dimension of the
longslit spectra in bins of 4 pixels (0.58 arcsecond). All velocities
presented are heliocentric and the line-of-sight zero point is defined
as the velocity of NGC\,1275, assumed to be 5265kms$^{-1}$
\citep{Ferruit}.  All distances referred to are projected distances.

\subsection{Northern filament}
\begin{figure*}
  \centering
\includegraphics[width=2\columnwidth]{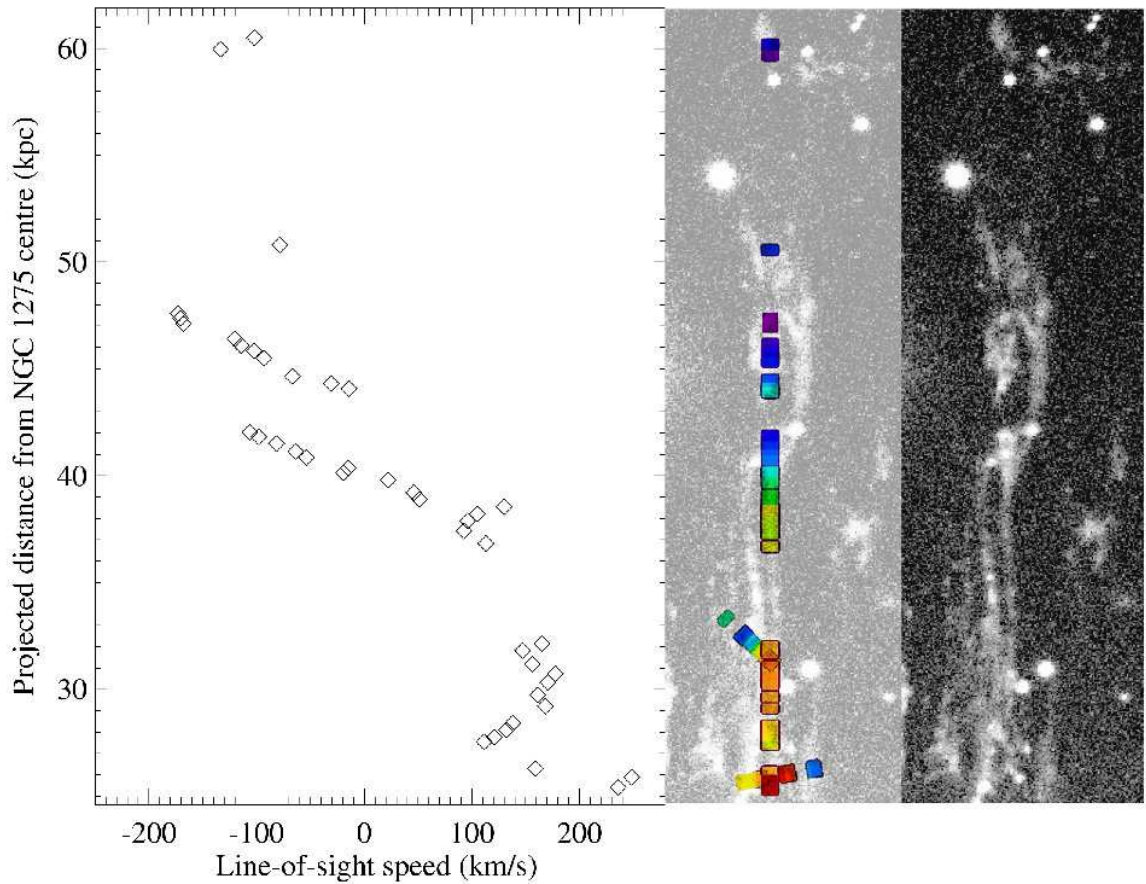} 
\caption{Line-of-sight velocities of the Northern filaments.
  Purple-blue indicate blueshifted emission, yellow-red indicate
  redshifted emission, whilst green has zero velocity relative to the
  central galaxy. Velocities from slit 1 (bottom) and slit 2 (top)
  which cut across the Northern filament are displayed in the image
  but not shown in graph. Background images are from the data of
  \citet{conselice}.\label{Northvela}}
\end{figure*}

The Northern filament is the dominant long ($\sim$60\,kpc), thin
($<$1\,kpc) structure stretching radially North-South, situated North
of NGC\,1275; slit 6 was positioned along this filament. The velocity
structure of the \ha\ and [N{\sc ii}] lines is shown in Figure
\ref{Northvela}. The filament appears extremely radial for the
majority of its length suggesting that the dominant direction of flow
is also radial. As this filament is the only structure detected so far
out from the galaxy it is extremely unlikely to be in projection with
another filament, and therefore we assume it to be a single structure.
It is improbable that we should be viewing an intrinsically-curved
filament as a linear one, so we assume that it is intrinsically
straight.

The filament has a kinematic North-South divide: the North displaying
a velocity blueshift by up to $-180$\kmps, whereas the South is
erratically redshifted.  The Northern half (above 37\,kpc) is clumpy
on scales of up to 5\,kpc in length. Each clump exhibits smooth
velocity gradients, although there are velocity discontinuities
between the clumps. It is possible that some of the velocity could be
due to the filament twisting as it falls (note the helical nature of
the lower parts of the filament, Fig. \ref{Northvela}). The Southern
part of this filament is split into two vertical segments: slit 6
covers only the western dimmer segment, whilst slits 1 and 2 slice
across both. The Western segment is very thin and redshifted.  The
Eastern segment appears thicker and clumpier, and slit 2 shows that
the emission is blueshifted.  Slit 1 shows that the very bottom of the
Eastern segment is redshifted.

The kinematic North-south divide indicates the lower part of the
filament is moving in the opposite direction to the upper part of the
filament: the filament is either being stretched or is collapsing,
depending on whether it is orientated toward or away from the
observer.  Half of this filament must be falling into the galaxy, and
the other half must be flowing away from the galaxy.  Thus we can
immediately rule out a model in which the filaments are smoothly
falling onto the galaxy below. The Doppler shifts alone do not enable
us to determine which end of the filament is inflowing or outflowing
since we then need to know the inclination. However, as part of the
filament must be flowing away from the galaxy, there must be a
mechanism for drawing gas away from the galaxy. 

We note that the Southern end of the three radial Northern filaments
coincides with a shock front seen in the X-ray images
\citep{Fabian2003b}. This front is due to the formation of the inner
Northern bubble around the radio source which is a cyclical process
taking place every $10^7\yr$ or so (as indicated by the presence of
the outer ghost bubbles).  If an expanding shock front destroys the
emission-line filaments, then the lower part of the Northern filaments
must previously have been at a larger radial distance in order to have
survived the shock emitted from the Northwest ghost bubble when it was
forming. Therefore the lower half of the Northern filament is
probably moving inward, whilst the upper segment is moving outward,
i.e.~the filament is likely to be stretching.

There is a depression in the thermal pressure just above the Northern
filament \citep{Fabian:2005:Perseus}, that could be a remnant of a
ghost bubble that has buoyantly risen from the central region. We can
now interpret the filament in the context of the rising bubble models
of \citet{BohringerM87,Churazov,Fabian2003,Reynolds}.  The radial
filament morphology traces the primary direction of flow therefore it
acts as a streamline. Part of the filament is flowing away from the
galaxy due to the uplift caused by the ghost bubble's buoyant rise
through the ICM, whilst the other half has been overcome by the
galaxy's gravity and is now flowing back. The pull from the bubbles
competes with gravity.

For a total filament length of 25\,kpc and a range in velocity of
400\kmps the dispersion time is 6$\times10^{7}$ years; if the filament
is at a small angle from the plane of the sky (as is likely due to its
large projected length) the velocity range may be much larger,
reducing the dispersion time.
\subsection{Northwest filaments and `horseshoe' feature}
\label{sec:horseshoe} 

To the Northwest of the galaxy lies an array of radial filaments, one
of which extends 30\,kpc from the nucleus and ends in a curved loop
that \citet{conselice} refer to as the `horseshoe' (detail in Fig.
\ref{hstfiladetail}). The loop is positioned underneath a ghost bubble
visible as a prominent depression in the X-ray image
\citep{Fabian2003b}.

\begin{figure*}
\centering
\includegraphics[width=1.0\columnwidth]{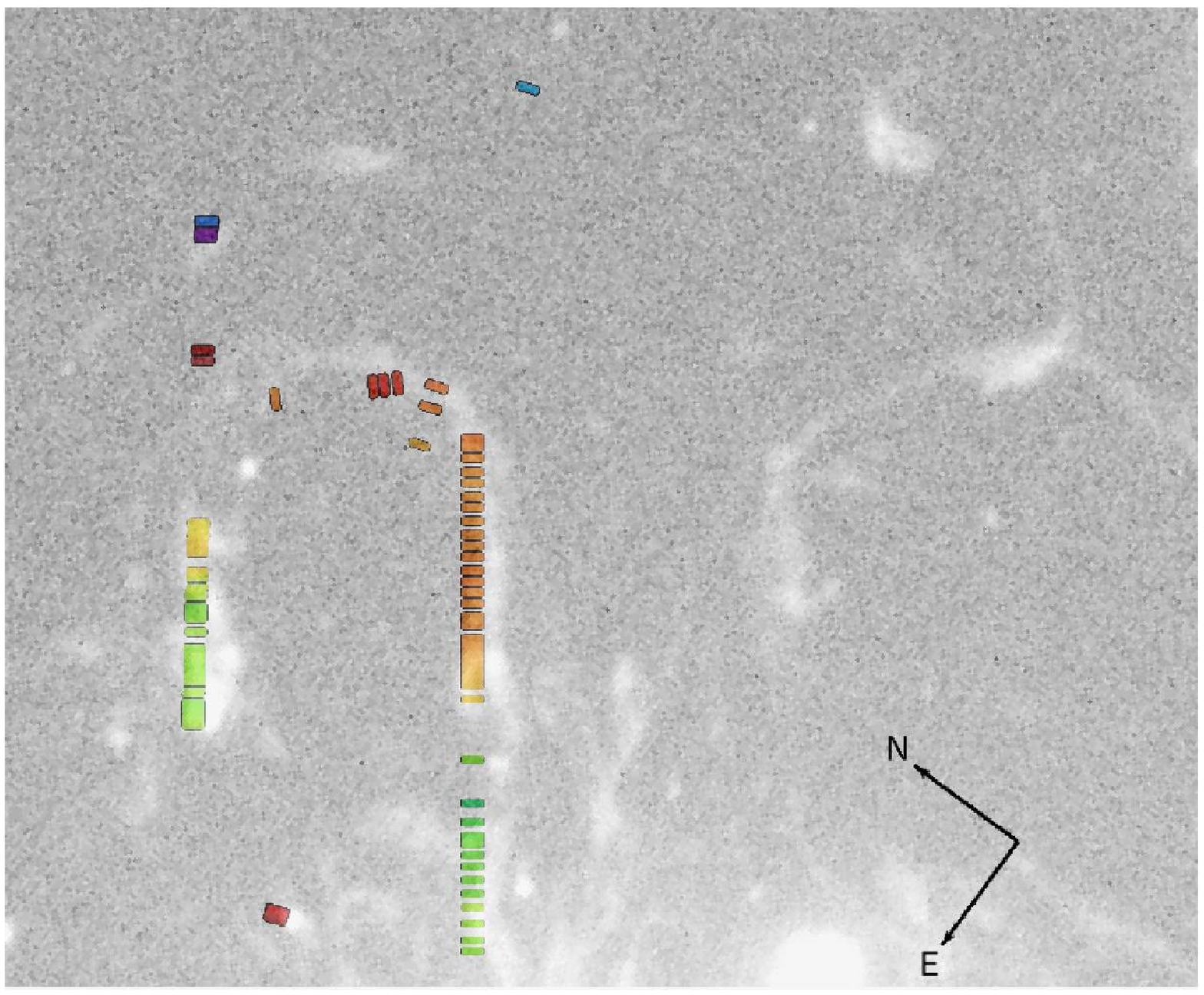}
\includegraphics[width=1.0\columnwidth]{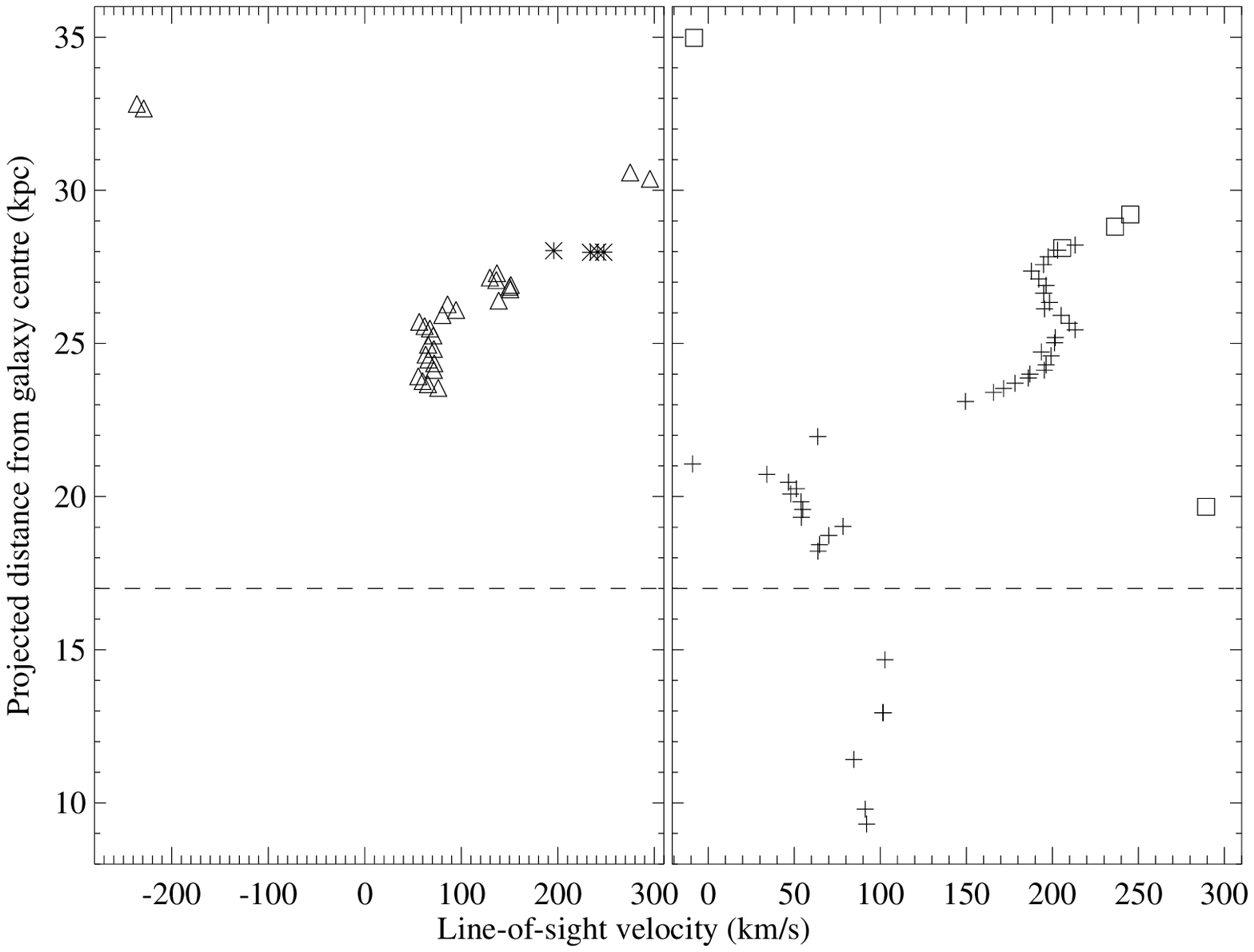}
\caption{ Left: Line-of-sight velocities along the `horseshoe' loop.
  Positive velocities are red, blue indicates negative velocities
  relative to galaxy. Right: Left panel shows velocities on the short
  straight of the loop covered by slit 3 (triangles) and along the top
  of the loop, covered by slit 2 (stars). Right panel shows data from
  the long straight on the right-hand side covered by slit 4 (crosses)
  and data from slit 5 that crossed through the loop (squares). Only
  data above dashed line is presented in image. Background image is
  from the data of \citep{conselice}.
  \label{movie}}
\end{figure*}

The morphology of these filaments has previously been noted to
resemble the flow underneath an air bubble rising in water
\citep{Fabian2003}. Figure \ref{movie} details the line-of-sight
velocities of these filaments. The long radial filament (Western part
of the `horseshoe' loop) begins with a redshifted line-of-sight
velocity of 95kms$^{-1}$ which remains fairly constant for 5\,kpc
until the line emission shifts Southwest beyond the slit for 3\,kpc,
to reappear at a distance of 18\,kpc from the galaxy with a velocity
of 60\,kms$^{-1}$.  The difference of 35\,kms$^{-1}$ between the two
sides of this gap cannot be unambiguously attributed to a change in
speed as a small change in orientation to the plane of our line of
sight could also produce the observed velocity deviations.

From 18\,kpc upward, this filament divides into two velocity
structures before curving into the loop which starts at 22\,kpc (see
right panel of Fig. \ref{movie}). These velocity structures have
smooth gradients with no small scale random deviations in excess of
the error (1-10\kmps\ depending on the line strength). The low
velocities, morphological structure of the filaments, and the
spherical cap appearance of the ghost bubble, indicate the filament
may be close to being in the plane of the sky.

The curved part of the `horseshoe' starts at a projected distance of
22\,kpc from the nucleus where the velocity increases rapidly to
200\,kms$^{-1}$, and remains steady over 4\,kpc. The top of the
loop has the highest velocity, peaking at 300\,kms$^{-1}$, then
slowing down to 200\,kms$^{-1}$ as the loop turns over. On the short
side of the loop the emission is still redshifted but the velocity
dies to 60\,km\ps\ in under 1\,kpc and remains steady
for the rest of the loop.  Above the loop is gas with a blueshifted
line-of-sight velocity of --230km\ps, a jump in velocity space of more
than 480km\ps\ over a projected distance of 3\,kpc.  Further above the
central axis of the bubble slit 5 cuts across some dim emission which
is also blueshifted. The gas in the loop and above the
loop surrounds the ghost bubble suggesting the gas above the loop is
part of the same structure as that in the loop and is not just a
projection effect.

The flow pattern qualitatively matches the simulations of
\citet{Reynolds} of a bubble rising through a viscous ICM: gas above
the bubble flows in the opposite direction to the gas below the
bubble, and the largest velocities occur near the central axis and
close to the bubble. If the flow of cool gas starts from the galaxy,
to stream up the long straight side, around the curve and down the
short straight side, we would expect the material in the short
straight side to be flowing in the opposite direction to the long
straight side, similar to that in an eddy, with blueshifted emission
underneath the bubble. Indeed the emission from the short straight
side is blueshifted relative to the rest of the loop: at a height of
25\,kpc from the nucleus, one side of the loop has a velocity of
200\,kms$^{-1}$ whereas the other side is at 60\,kms$^{-1}$. 

The agreement of the velocities, morphology and the clear ghost bubble
visible in the X-ray images of the Perseus cluster
\citep{Fabian2003b}, suggest the most likely dynamical model is one in
which the loop and radial filament is flowing out of the galaxy, with
the filament slightly orientated away from us. As the bubble has risen
(possibly with velocities of 700km\ps \citealt{Fabian2003}), it has
dragged up cool material from the galaxy producing the filamentary
structure we observe.

\subsection{Tangential filament}

\begin{figure*}
  \centering
\includegraphics[width=2\columnwidth]{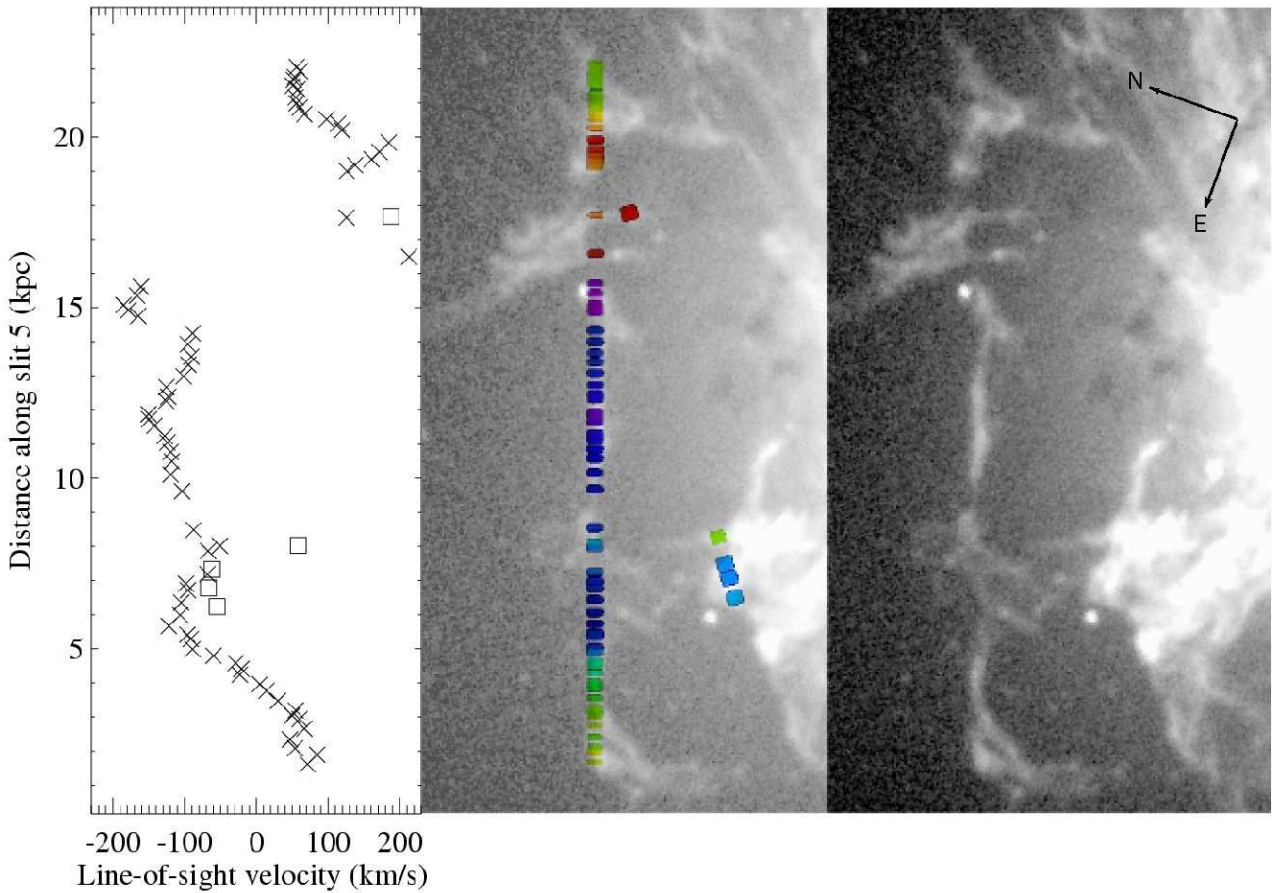}
\caption{Line-of-sight velocities of the tangential filament running
  along the Northeast of NGC\,1275. Purple-blue indicate blueshifted
  emission, yellow-red is redshifted emission and green has zero
  velocity relative to the central galaxy. Data from slit 5 is
  presented in the graph with crosses. Slit 3 also covered some nearby
  regions which have been marked by the square symbols. Luminous
  radial filaments extending from the galaxy to the tangential
  filament appear at the bottom right of the image. Background image
  is from the data of \citet{conselice}. \label{tang}}
\end{figure*}
Fig. \ref{tang} shows the line-of-sight velocities of the filament
that appears to run tangentially along the North of the galaxy. Slit 3
also covered line emitting regions situated between the tangential
filament and the galaxy whose velocities have been added into Figure
\ref{tang} as square symbols. 

The velocity structure along 14.5\,kpc varies without large velocity
jumps so the Eastern section of the filament appears to be a coherent
structure. The emission is blueshifted to a similar velocity along
$\sim$10\,kpc of its length then smoothly decreases in speed at the
most Eastern end. Small scale deviation from the large scale trends
are seen in excess of the error ($\sim$1--10\kmps\ depending on the
line intensity), suggesting some small scale random variations in
velocity. Beyond 14.5\,kpc the emission first jumps by --80\,km\ps\ in
velocity then jumps again by almost +400\,km\ps.  The Western section
of this filament is not coherent in velocity space, and the morphology
suggests that slit 5 is slicing across radial filaments that extend to
the Northwest, in the same direction as the `horseshoe' feature. The
Eastern section is a puzzling structure since it is tangential, unlike
the majority of the filaments. Interpretation of this region is
complicated by the presence of the Northern radio lobe and complex
X-ray emission.

\section{Filament line-widths and velocities}

\begin{figure}
\centering
\includegraphics[width=1.0\columnwidth]{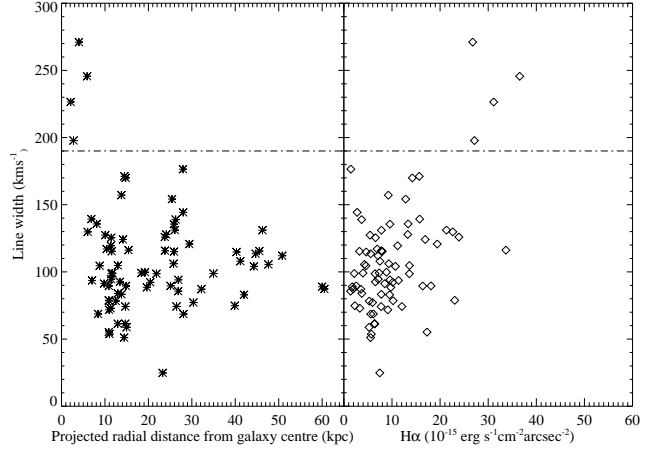}
\caption{{\bf Left}: Radial projection of emission line widths in
  kms$^{-1}$ based on the \ha\ and [N{\sc ii}] lines. {\bf Right}: Line
  width verses \ha\ surface brightness. Most points with large line-widths
  ($>$200\kmps) are within the central few kpc and have large \ha\
  surface brightness suggesting they may be overlapping filaments.  }
\label{linewidth}
\end{figure}

\begin{figure}
 \centering
 \includegraphics[height=1.0\columnwidth, angle=-90]{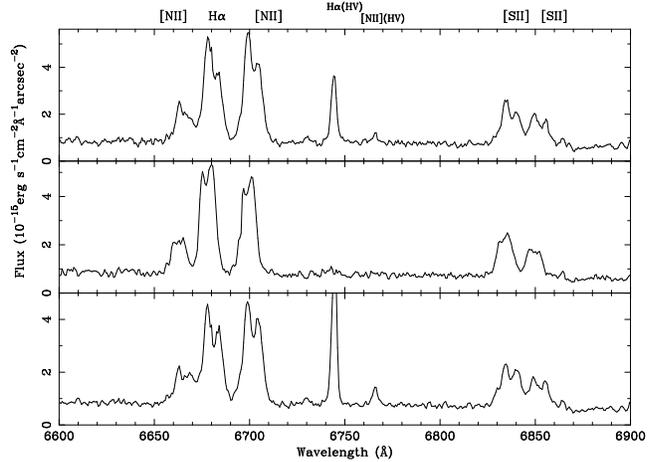}
 \caption{Examples of central clumps which show double-peaked line
   emission. These regions are all centrally located. HV denotes
   emission from the high-velocity system which lies infront of
   NGC\,1275.}
   \label{doublexamples} 
 \end{figure}

The FWHM (full width at half maximum) of the instrumental profile is
85\kmps\ (2.85 pixels) at \ha\ determined from nearby sky lines. Most
of the material has FWHM line-widths of 50-160\,kms$^{-1}$ after
correcting for the instrumental broadening (Fig.  \ref{linewidth}),
much greater than the thermal width of hydrogen gas at 10,000\,K
($\sim$20\,kms$^{-1}$). If the filaments represent an inflow of gas as
predicted by early models of cooling flows \citep{FNC} we would expect
an anti-correlation between line-widths and radial distance from the
nucleus and extremely large ($\sim$500--1000\,kms$^{-1}$) central
line-widths \citep{Heckman}.  We observe a few points within 10\,kpc
that have large ($>$200\,kms$^{-1}$) line-widths. It is within 10\,kpc
of the galactic centre that the density of the line-emitting filaments
increases greatly and there are many regions where the spectra display
double peaked lines indicating that the line-of-sight crosses at least
two clumps of line-emitting material which have different kinematics.
It is not necessary that these clumps be physically close, therefore
they do not imply small-scale velocity deviations along a single
filament as observed in slit 5.  Examples of such regions are shown in
Fig.  \ref{doublexamples}, and were either resolved into two sets of
lines or removed from the dataset if the result was ambiguous.
However, it is likely that some of these central regions would have
clumps with similar line-of-sight velocities and result in spectra
with a single wide peak. Most regions with large line-widths also have
a large \ha\ surface brightness (Fig.  \ref{linewidth}), therefore it
is probable that the spectra from these inner regions are caused by
filaments overlapping in the same line-of-sight with slightly
different velocities. Beyond the inner 10\,kpc, the line-widths are
uniformly 2--8 times the thermal width of gas at 10,000\,K.  Some
radially extending filaments exhibit similar line-widths along their
whole length (see Fig.  \ref{widths4} in section
\ref{sec:Lineratios}).  Therefore the line-widths provide no evidence
to suggest the nebula flows into the galaxy.

No line-of-sight velocity greater than 350\,km\ps was detected. This
work primarily probes the outer filaments, which are likely to have
small angles from the plane of the sky due to their large projected
distance from the galaxy, and therefore are not expected to have large
line-of-sight velocities. \citet{Cigan} who probe the central regions
as well as the outer filaments find no velocities greater than
450\,km\ps.  In section \ref{sec:horseshoe} we argue that the
`horseshoe' feature and the Northwest filaments covered by slit 4 are
very close to being in the plane of the sky.  Therefore the observed
line-of-sight velocity of 200\,km\ps, at the top of the `horseshoe'
loop could transform to a velocity much greater than 700\,km\ps
(expected if tilted by a conservative 75$^{\circ}$ from the
line-of-sight).  This is far beyond what is observed in the rest of
the dataset. It is possible that the two inner radio lobes have pushed
or destroyed the filaments pointing toward our line-of-sight.


\begin{figure*}
 \centering
 \includegraphics[width=1.5\columnwidth]{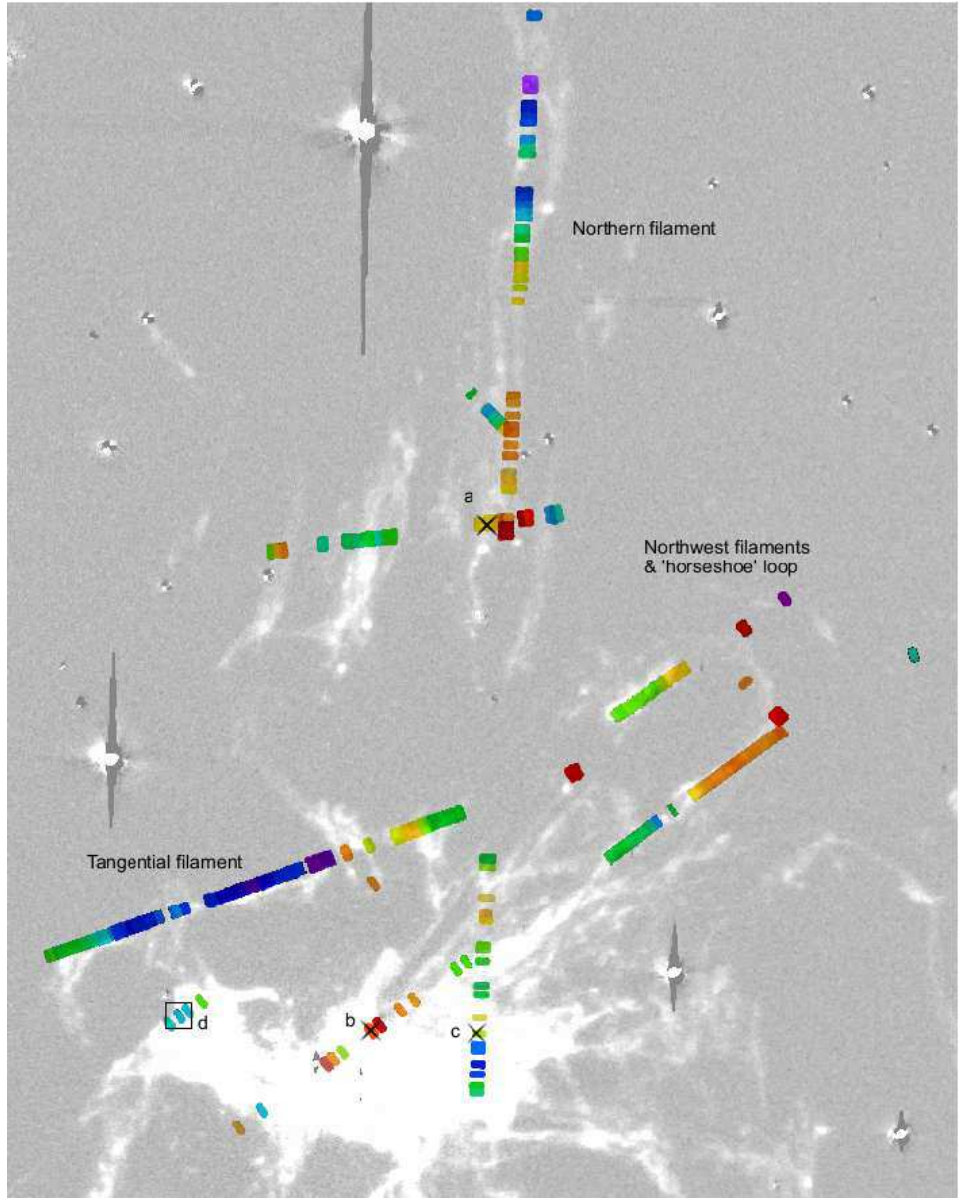}
 \caption{All velocities and regions from which spectra are shown in
   this work are marked. The region marked by a square and labelled
   {\bf d} is where the typical spectrum from Fig. \ref{spectrum} is
   extracted. The regions marked by a cross and labelled {\bf a}, {\bf
     b} and {\bf c} are the regions where [O{\sc iii}]$\lambda$5007 is
   detected; the spectra are displayed in section \ref{sec:OIII}.  The
   regions which exhibit double peaked emission come from the central
   areas covered by slits 4 and 6 and are located near the labels {\bf
     b} and {\bf c}. The background image comes from the data of
   \citet{conselice}.}
   \label{ALL} 
 \end{figure*}

 \section{Spectral features}
\label{sec:ratios} 
 \begin{figure*}
  \includegraphics[height=2.0\columnwidth,angle=-90]{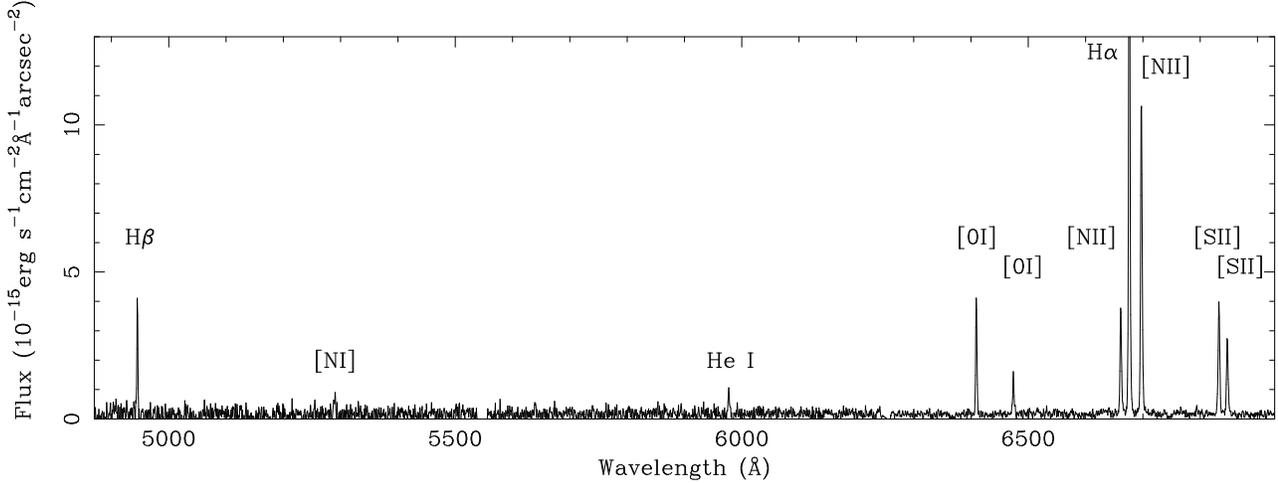}
  \caption{A typical spectrum showing the (low-velocity system)
    redshifted lines of [O{\sc i}], [N{\sc ii}], \ha, \hb, and [S{\sc
      ii}].  He{\sc i} and [N{\sc i}] only appear in very luminous
    regions. The full strength of \ha\ is not shown.}
  \label{spectrum} 
\end{figure*} 
\begin{figure}
  \includegraphics[width=1\columnwidth]{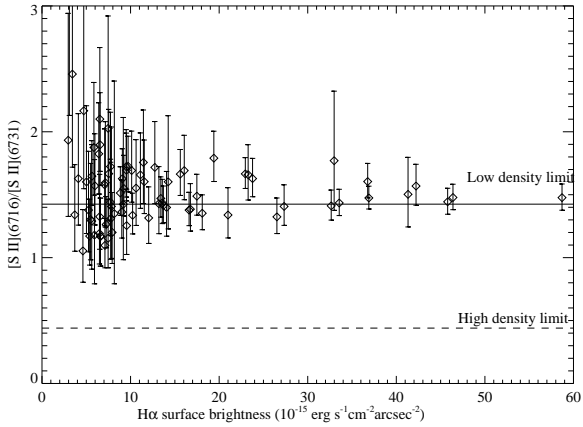}
  \caption{[S{\sc ii}]$\lambda$6716/[S{\sc ii}]$\lambda$6731 shows the
    gas is below the low density limit of $10^{2}$cm$^{-3}$}
  \label{ss} 
\end{figure} 
\begin{table*}
  \begin{tabular}{|lllllllll|}\hline
    & \ha\ surface brightness &[N{\sc ii}]$\lambda$6584/ \ha\ &  [S{\sc ii}]$\lambda$6716/ \ha\ & [O{\sc i}]$\lambda$6300/\ha\ & [S{\sc ii}]$\lambda$6716/ &\ha /\hb & He{\sc i}$\lambda$5876/\hb\ &[N{\sc i}]$\lambda$5199/\hb\\
    & ($10^{-15}$\ergpcmsqpspsqarcsec) & & & & [S{\sc ii}]$\lambda$6731\\ \hline
    Typical &  $<$15& 0.3-1.5&0.14-0.54&0.1-0.3&1.4&3.5-7&0.2-0.3 &0.4  \\
    values& & & & & &\\ \hline 
\end{tabular} 
\caption{Range of typical line intensity ratios within the extended nebula} 
\label{filamentsummery}
\end{table*}

\label{sec:Lineratios} 
Our data probe the outer nebula that extends beyond 10\,kpc in detail
for the first time. All intensity line ratios and line intensities are
determined from binning the spatial dimension of the longslit spectra
in bins of 6 pixels (0.87 arcsecond). A typical spectrum is shown in
Fig.  \ref{spectrum} and line intensity ratios are summarised in Table
\ref{filamentsummery}.  All spectra exhibit low-excitation features:
high [N{\sc ii}], [O{\sc i}] and [S{\sc ii}] compared to \ha. The
appearance of strong [O{\sc i}] and [S{\sc ii}] confirm we are
observing an ionization bound nebula as these species are observed
from the partially ionized region beyond the classical H{\sc ii}
region. There is a whole body of material on the optical line ratios
from the central regions of this object and the reader is referred to
\citet{Kent,Heckman,SabraShieldsFlip,conselice} and references
therein. The low-ionization line emission extends from the central
regions to the outer filaments. In the following sections we
concentrate on the features illuminated by the study of the outer
filaments. The \ha\ intensities range between
2--57$\times10^{-15}$\ergpcmsqpspsqarcsec, with the majority less than
15$\times10^{-15}$\ergpcmsqpspsqarcsec. [N{\sc ii}]$\lambda$6584/\ha\
ratios tend to be higher near the nucleus.  The line intensity ratio
[S{\sc ii}]$\lambda$6716/[S{\sc ii}]$\lambda$6731 shown in Fig.
\ref{ss} confirms the gas is in the low density limit, therefore is
less than $10^{2}$cm$^{-3}$ \citep{Osterbrock}, which agrees with
measurements based on \ha\ surface brightness indicating the electron
density is greater than 10\,cm$^{-3}$ \citep{conselice}.

\subsection{Radial variation in emission-line ratios}
\begin{figure}
  \includegraphics[width=1\columnwidth]{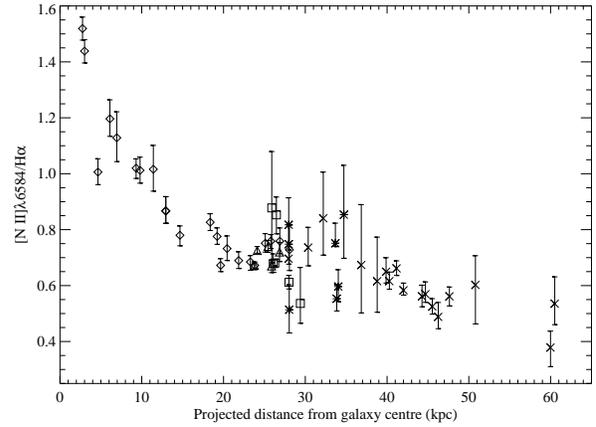}
  \caption{[N{\sc ii}]$\lambda$6584/\ha\ as a function of projected
    distance from the galactic centre. Data from slit 1 (squares),
    slit2 (stars), slit3 (triangles), slit 4 (crosses) and slit 6
    (diamonds). 1$\sigma$ error bars from error of line-fitting.}
\label{nhslit4} 
\end{figure}

\begin{figure}
  \includegraphics[width=1\columnwidth]{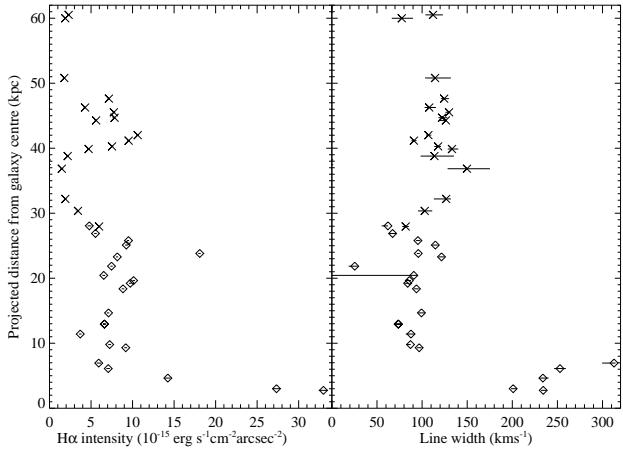}
  \caption{Line-width and \ha\ intensity shown as a function of
    distance from the galactic centre. Data taken from slit 4
    (crosses) and slit 6 (diamonds). 1$\sigma$ error bars from error
    of line-fitting.}
\label{widths4} 
\end{figure}

The outer radial filaments to the North and Northwest show a clear
correlation between the \nii/\ha\ intensity ratio and projected
distance from the central galaxy.  Fig.~\ref{nhslit4} shows the
variation of [N{\sc ii}]$\lambda$6584/ \ha\ along slits 1,2,3,4 and 6,
which covers the 30\,kpc radial filament that extends from the galaxy
to the `horseshoe' feature and the Northern filaments.  Data from slit
5, the Eastern regions covered by slit 3, and the lower 15\,kpc
section covered by slit 6 were not included as the nebula morphology
suggests these regions cut across many radial filaments
(Fig.~\ref{ALL}), hence the projected distance may not be a good
approximation to the actual distance. Line-emitting regions within
4\,kpc of the nucleus covered by slit 4 have been removed from the
dataset as the large line-widths and \ha\ luminosity
(Fig.~\ref{widths4}) suggest these sections are overlapping filaments.
A clear radial gradient in the [N{\sc ii}]$\lambda$6584/ \ha\ is
observed. The slope is steeper up to 30\,kpc than between 30--60\,kpc
which may be due to a projection effect. The Northwest filaments may
have a larger inclination from the plane of the sky than the Northern
filaments, therefore the projected distances below 30\,kpc may
correspond to a larger true distance compared to the datapoints from
the Northern filaments which lie between 30--60\,kpc. Such a gradient
has also been observed in M87, the central galaxy of the Virgo cluster
(Sarzi, M; private communication).


The \ha\ surface brightness remains constant along the filaments
(Fig.~\ref{widths4}), which rules out a central ionizing source which
would produce a gradient in the \ha\ luminosity \citep{RodAndy}.  If
the filaments were excited by shocks we would observe a gradient in
the line-width as well as in the \nii/\ha\ intensity ratio. Models of
fast shocks by \citet{Dopita} imply a range of FWHM of \ha\ between
150--300\kmps should be observed, therefore our data rule out fast
shocks, but do not rule out shocks of moderate speed as modelling of
\citet{Shull} imply a very steep gradient between a FWHM of
70--130\kmps would be observed. Our data show a great deal of scatter.

\begin{figure}
  \includegraphics[width=1\columnwidth]{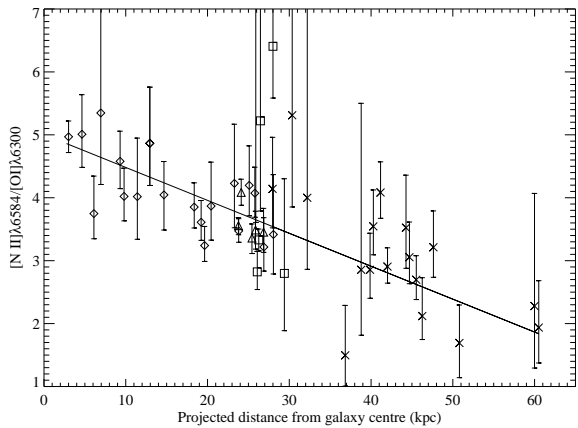}
  \caption{Data from slit 1 (squares), slit2 (stars), slit3
    (triangles), slit 4 (crosses) and slit 6 (diamonds) showing
    variation of [N{\sc ii}]$\lambda$6584/[O{\sc i}]$\lambda$6300 with
    distance from nucleus. The \nii\ line intensity increases more
    than the [O{\sc i}]$\lambda$6300 line intensity. Lines indicates
    best fit to the data. 1$\sigma$ error bars from error of
    line-fitting.}
  \label{n2o1dist}
\end{figure}

[N{\sc ii}] emission is produced by collisional excitation with
electrons so the intensity of the line is a measure of the heat within
the gas. The \ha\ line intensity measures the ionization rate within
the gas, so the increasing [N{\sc ii}]$\lambda$6584/\ha\ line
intensity ratio may indicate that more heating per Hydrogen ionization
occurs within the gas close to the galaxy compared with the gas
further out.  Collisional excitation can be more efficient in dense
environments, so this gradient could mirror the pressure variation of
the surrounding ICM.  The soft X-ray filaments detected by
\citet{Fabian2003} spatially coincide with the optical filaments,
indicating an energy exchange occurs between the two gas phases. The
[N{\sc ii}]$\lambda$6584/\ha\ gradient may be a further symptom of the
interaction between the warm optical gas and the hot ICM.
Alternatively this trend may be attributed to an extra heating
mechanism originating from the central parts of NGC\,1275. Increasing
[N{\sc ii}]$\lambda$6584/\ha\ and [S{\sc ii}]$\lambda$6717/\ha\
intensity ratios have been observed extending above the plane of the
Milky Way and of other spiral galaxies \citep{Miller} and is
attributed to an extra heating mechanism \citep{Elwert}.

The [N{\sc ii}]$\lambda$6584/\ha\ ratio is also sensitive to
variations in metallicity. An increase in the nitrogen abundance would
increase the [N{\sc ii}]$\lambda$6584/\ha\ ratio, however, as oxygen
is an efficient coolant, an increase in the oxygen abundance would
decrease the gas temperature and reduce the [N{\sc
  ii}]$\lambda$6584/\ha\ ratio. Therefore the \nii/\ha\ line intensity
ratio is sensitive to the nitrogen/oxygen abundance in the nebula.
Fig.~\ref{n2o1dist} plots the [N{\sc ii}]$\lambda$6584/[O{\sc
  i}]$\lambda$6300 ratio against projected distance for the same
regions plotted in Fig.~\ref{nhslit4}. A small radial variation may be
present, however the errors are too large to be conclusive. How
metallicity variations affect the [N{\sc ii}]$\lambda$6584/\ha\ or
[N{\sc ii}]$\lambda$6584/[O{\sc i}]$\lambda$6300 ratio depends on the
excitation mechanism. Since both the metallicity and excitation source
are currently unknown, we cannot determine whether the radial
variation in line ratios is due to metallicity changes or an
excitation process.

\subsection{\ha /\hb ratio}
The Balmer decrement for the filaments of NGC\,1275 is high,
\citet{Kent} place the value at 4.77 which corresponds to {\it E}({\it
  B-V})=0.43 under the assumption of case B recombination theory. Our
data of the extended filaments, that are distinct from the
high-velocity system, also have a high \ha/\hb\ ratio, ranging over
3--7.  We find \ha/\hb$\sim3$ for the high velocity system in
agreement with 3.15 from \citet{Kent}.  Therefore the decrement is
caused by a process occurring in the filaments themselves, such as the
photoionization mechanism suggested by \citet{Donahue91}, involving
EUV and soft X-rays would naturally lead to high \ha/\hb\ ratios, or a
great deal of dust either within the filaments or between the
infalling galaxy and the low-velocity filamentary nebula.  There is
clearly a lot of dust associated with the high-velocity system as it
appears in absorption in HST and X-ray images \citep{Gillmon}, however
it is confined to the Northwest of NGC\,1275.  The dynamical model we
present in section \ref{sec:kinematics} implies the filaments are
drawn out of the galaxy, therefore it is possible that the filaments
are dusty. By contrast, if the filaments had condensed out of the ICM
we would not expect them to immediately contain dust \citep{FNC}.

\subsection{Faint line-emission} 
\subsubsection{[O{\sc iii}] line-emission}
\label{sec:OIII}
The [O{\sc iii}]$\lambda$4959 and $\lambda$5007 emission lines are not
detected in most of the longslit spectra in agreement with
observations of \citet{Meaburn}. These lines are typically found in
both high and low excitation spectra, and are commonly produced by hot
stars, planetary nebulae, and active galactic nuclei. Studies of the
inner nebula often detected [O{\sc iii}]$\lambda$4959, $\lambda$5007
emission (e.g. \citealt{Kent,SabraShieldsFlip}). Since oxygen is
clearly present throughout the whole nebula given the strong [O{\sc
  i}]$\lambda$6300 emission line, it is probable that there is an
additional excitation mechanism acting on the inner nebula which
produces the [O{\sc iii}]$\lambda$4959, $\lambda$5007 line-emission.
The ionization mechanism in the outer nebula may not be hard enough to
produce O$^{++}$ which requires 54.93eV (similar to He{\sc ii} which
has not been detected in the NGC\,1275 nebula). It is unlikely that
the lack of [O{\sc iii}]$\lambda$4959, $\lambda$5007 emission is due
to a very high oxygen abundance in the outer nebula compared to the
inner nebula. A high oxygen abundance would cool the nebula through
far-infrared lines at 52$\mu$m and 88$\mu$m which would result in
reducing the optical [O{\sc iii}] emission. It is even more improbable
that the [O{\sc iii}] lines are suppressed by collisional
de-excitation which starts at a density of $10^{6}$cm$^{-3}$, as the
density measured from the [S{\sc ii}] lines is below this value by at
least 3 orders of magnitude.

\begin{figure}
\includegraphics[height=1\columnwidth, angle=-90]{Fig16a.ps}
\includegraphics[width=1\columnwidth]{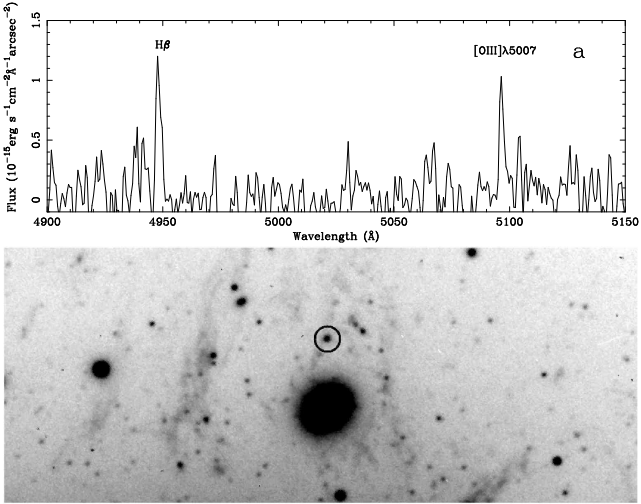}
\caption{Spectrum of highlighted region showing clear [O{\sc iii}]
  line. Black circle highlights a bright knot from which the [O{\sc
    iii}] is detected.}
\label{spectrumOIII}
\end{figure}

[O{\sc iii}]$\lambda$5007 emission is detected in only three regions.
The clearest detection occurs in slit 1 in the region marked by a
circle in Fig.~\ref{spectrumOIII} with [O{\sc
  iii}]$\lambda$5007/\hb=0.78. This region also has continuum emission
although all other features of the spectrum are fairly typical of the
rest of the filaments (see Table\ref{OIIItable} for details).

\begin{figure}
\includegraphics[height=1.05\columnwidth, angle=-90]{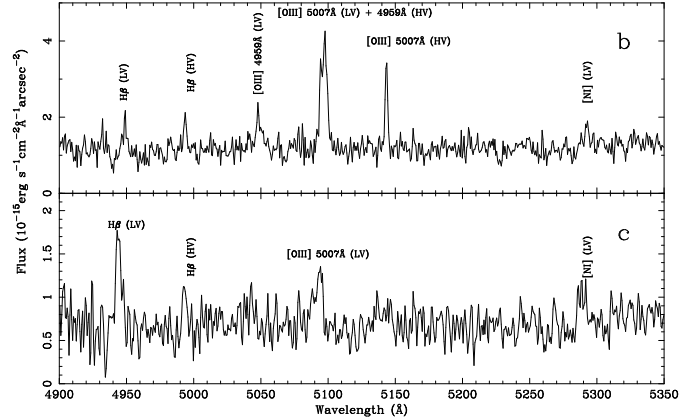}
\caption{Spectra of complex regions showing [O{\sc iii}] lines. HV and
  LV denotes the high-velocity system and the low-velocity system
  respectively. The high-velocity system is an infalling galaxy
  infront of NGC\,1275. The low-velocity system is the filamentary
  nebula surrounding NGC\,1275.}
\label{spectrumOIII2}
\end{figure}

The other two spectra which have [O{\sc iii}] lines are complex as
both lie close to the galaxy centre and overlap the high-velocity
system (Fig.~\ref{spectrumOIII2}). The [O{\sc iii}]$\lambda$5007 line
of the low-velocity system overlaps the [O{\sc iii}]$\lambda$4959 line
from the high-velocity system.  This is particularly bad in spectrum
{\it b} (Fig.~\ref{spectrumOIII2}) as the [O{\sc iii}]$\lambda$5007
line from the high-velocity system is bright. Therefore the [O{\sc
  iii}]$\lambda$5007 emission from the low-velocity system has been
estimated from the $\lambda$4959 line using an intensity ratio of 2.98
\citep{Storey}.  Whilst the spectrum {\it c} has similar features to
the clear detection in spectrum {\it a}, spectrum {\it b} has very
different ratios. The spectral features are summarised in Table
\ref{OIIItable}.  It should be noted that spectra {\it b} and {\it c}
also exhibit [N{\sc i}]$\lambda$5199 emission which shall be discussed
in the following section.  If the ratio was set at 0.7 (as in spectrum
a and b), the [O{\sc iii}] 5007\AA\ emission should be detectable in
the majority of the longslits, therefore these regions are an
exception in comparison to the rest of the studied nebula.
\begin{table}
\begin{tabular}{|lllll|}
& [O{\sc iii}]$\lambda$5007/ \hb\ &  [N{\sc ii}]$\lambda$6584/\ha\  & [S{\sc ii}]$\lambda$6716/ & \ha\
 surface\\ 
&   &  & [S{\sc ii}]$\lambda$6731 & brightness\\  \hline
a&0.78&0.61&1.32&12.0\\ 
b& 3.5&1.44&1.41&27.3\\ 
c&0.61&0.87&1.41&32.7\\ \hline
\end{tabular}
\caption{Line ratios of regions in which [O{\sc iii}] 5007\AA\ emission is detected. \ha\
 surface brightness in units of ($10^{-15}$\ergpcmsqpspsqarcsec). }
\label{OIIItable}
\end{table}

\subsubsection{He {\sc i} and [N{\sc i}] line-emission}
He{\sc i}$\lambda$5876 and [N{\sc i}]$\lambda$5199 emission lines are
commonly detected in the NGC\,1275 nebula where the \ha\ surface
brightness exceeds 30$\times10^{-15}$\ergpcmsqpspsqarcsec. [N{\sc i}]
was only detected within 8\,kpc of the nucleus, but there was one
detection of He{\sc i} at a distance of 24\,kpc from the nucleus, in
the tip of the horseshoe loop. In total [N{\sc i}] was detected in 7
regions whilst He{\sc i} was detected in 5 regions; only the region at
the tip of the horseshoe loop had detectable He{\sc i} yet no clear
[N{\sc i}]. Given the low He{\sc i}/\hb\ and [N{\sc i}]/\hb\ ratios of
$\sim$0.25 and $\sim$0.4 respectively, it is possible that all the
filaments emit through these lines, but the signal-to-noise of the
observations is not high enough to detect them. The combination of
[O{\sc i}], [N{\sc i}] and He{\sc i} emission suggests that the
filaments have a layered structure such as observed in the Crab nebula
\citep{Sankrit}. Warm molecular gas has been found at the same
location as the filaments \citep{Hatch} which implies there is a cool
(2000\,K) component associated with the ionized gas.

\section{Conclusion}
The kinematic data presented in section \ref{sec:kinematics} rules out
any dynamical models of purely infalling filaments.  The low
velocities and line-widths presented here argue strongly against
inflow, as one would expect both these parameters to rise sharply
toward the centre of the nebula \citep{Heckman,Donahue91}. The most
conclusive evidence lies in the velocity structure of the Northern and
Northwest filaments.  The lower half of the Northern filament is
redshifted,whilst the upper section is blueshifted, thus the upper
section of the filament is moving in the opposite direction to the
material below: part of the filament is flowing away from the galaxy,
whilst the other part is flowing into the galaxy.  In order to explain
the outflow we appeal to the models of \citet{BohringerM87,Churazov}
and \citet{Fabian2003} in which the radio emitting plasma from the AGN
forms bubbles in the ICM, which detach to become buoyant and rise,
dragging cool material from the galaxy below. The Northwest filaments
lie directly underneath a ghost bubble \citep{Fabian2003}. In addition
to the morphological resemblance noted by \citet{Fabian2003}, the
kinematic signature of these filaments matches simulations of gas flow
under a buoyantly rising bubble \citep{Reynolds}, including details
such as gas above the bubble moving in the opposite direction to the
filaments below.

The data suggests the filaments are outflowing and therefore their
origins lie within the galaxy. NGC\,1275 contains a large reservoir of
cold molecular gas \citep{Krabbe,Donahue} that can fuel these
filaments. As we observe part of the Northern filament falling back
into the galaxy it is possible that the fate of all filaments lie in an
eventual return to the galaxy.  However, if the filament falls back in
segments as observed in the Northern filament, this would stretch and
possible narrow the filament, making them more susceptible to
evaporation by the ICM.  The detection of [O{\sc i}] and [N{\sc i}]
indicates the presence of warm atomic hydrogen, and warm molecular
hydrogen has been found in the outer filaments \citep{Hatch}. It is
possible the filaments holds a significant amount of cooler, so-far
undetected gas.  In the manner proposed above, the central galaxy can
efficiently lose mass and pollute the ICM with metals.

We report a radial variation of the [N{\sc ii}]$\lambda$6584/\ha\
ratio, indicating either progressive hardening of the excitation
mechanism close to NGC\,1275, or a variation in the nitrogen/oxygen
abundance. 

Although NGC\,1275 is surrounded by numerous stellar clusters, we have
presented details of HST images which show there is no preferential
association with the optical filaments (Figs. \ref{hstloops} and
\ref{hstfiladetail}).  As [O{\sc iii}] emission (commonly found in
H{\sc ii} regions) is also lacking in the filaments, it is unlikely to
be powered by or the birthplace of hot young stars. The central region
exhibits [O{\sc iii}]$\lambda$5007 line emission, in contrast to the
outer nebula, therefore an additional hard excitation source may be
influential in the central region.

\section*{Acknowledgements}
NAH and RMJ acknowledge support from PPARC and ACF and CSC thank the
Royal Society for support.\\
Data from sections \ref{sec:continuum}, \ref{sec:kinematics} and
\ref{sec:ratios} based on observations obtained at the Gemini
Observatory, which is operated by the Association of Universities for
Research in Astronomy, Inc., under a cooperative agreement with the
NSF on behalf of the Gemini partnership: the National Science
Foundation (United States), the Particle Physics and Astronomy
Research Council (United Kingdom), the National Research Council
(Canada), CONICYT (Chile), the Australian Research Council
(Australia), CNPq (Brazil) and CONICET (Argentina).  Images from section
\ref{sec:HST} based on observations made with the NASA/ESA Hubble
Space Telescope, obtained from the data archive at the Space Telescope
Institute. STScI is operated by the association of Universities for
Research in Astronomy, Inc. under the NASA contract NAS 5-26555.

\bibliographystyle{mnras} \bibliography{mn-jour,wotable}

 \end{document}